\documentclass[onecolumn]{aastex63}

\shorttitle{Multi-Wavelength TRGB}
\shortauthors{Madore, Freedman \& Owens}
\graphicspath{{./}{figures/}}

\begin{document}

\title{\bf Astrophysical Distance Scale ~VII: \\ A Self-Consistent, Multi-Wavelength Calibration of the Slopes and Relative Zero Points for the \\ Run of Luminosity with Color of Stars Defining the Tip of the Red Giant Branch}
\author[0000-0002-1576-1676]{\bf Barry F. Madore} 
\affil{The Observatories, Carnegie
Institution for Science, 813 Santa Barbara St., 
Pasadena, CA ~~91101, USA}
\affil{Department of Astronomy \& Astrophysics, University of Chicago, 5640 S. Ellis Ave., Chicago, IL 60637, USA}
\email{barry.f.madore@gmail.com}
\author[0000-0003-3431-9135]{\bf Wendy~L.~Freedman}
\affil{Department of Astronomy \& Astrophysics, University of Chicago, 5640 S. Ellis Ave., Chicago, IL 60637, USA}
\affil{Kavli Institute for Cosmological Physics, University of Chicago, 5640 S. Ellis Ave., Chicago, IL 60637, USA}
\email{wfreedman@uchicago.edu}
\author[0000-0003-3339-8820]{\bf Kayla Owens}
\affil{Department of Astronomy \& Astrophysics, University of Chicago, 5640 S. Ellis Ave., Chicago, IL 60637, USA}
\affil{Kavli Institute for Cosmological Physics, University of Chicago, 5640 S. Ellis Ave., Chicago, IL 60637, USA}
\email{kaowens@uchicago.edu}

\begin{abstract} 
Given the recent successful launch of the \textit{James Webb Space Telescope}, determining robust calibrations of the slopes and absolute magnitudes of the near- to mid-infrared Tip of the Red Giant Branch (TRGB) will be essential to measuring precise extragalactic distances via this method. Using ground-based data of the Large Magellanic Cloud from the Magellanic Clouds Photometric Survey along with near-infrared (NIR) data from 2MASS and mid-infrared (MIR) data collected as a part of the SAGE survey using the \textit{Spitzer Space Telescope}, we present slopes and zero-points for the TRGB in the optical (VI), NIR (JHK) and MIR ([3.6] \& [4.5]) bandpasses. These calibrations utilize stars +0.3$\pm$0.1 mag below the tip, providing a substantial statistical improvement over previous calibrations which only used the sample of stars narrowly encompassing the tip. 
\end{abstract}




\section{Introduction}

With the successful launch of the {\it James Webb Space Telescope} (JWST) and the continuing uncertainty in the local value of the Hubble constant (Freedman 2021), the Tip of the Red Giant Branch (TRGB) Method has become even more important in the endeavor to calibrate the local expansion rate of the Universe. 
Given the light-gathering power and unequalled point spread function (PSF) resolution of JWST in the near-infrared (NIR) and mid-infrared (MIR), it now seems an appropriate time to enumerate and quantify the advantages, as well as to assess any disadvantages, in application of the TRGB method into the infrared.  In addition, it is useful to explore optimal filter combinations and observing strategies for calibrating those relations.

In the following we present a fully empirical approach to robustly and self-consistently calibrating optical (VI), NIR (JHK) and for the first time, the MIR (3.6 and 4.5 $\mu$m) TRGB relations spanning the wavelengths covered most sensitively by the NIRSS and NIRCAM imagers on JWST. 
The calibrations presented here will be indicative of the slopes and zero points expected for closely similar filters available on those cameras. 
Ultimately those flight magnitudes will have to be tied into the ground-based calibrations presented here using short exposures on very nearby TRGB fields as already observed from the ground and from space (by HST). 
Two obvious ``bridging galaxies'' are WLM  (Wolf-Lundmark-Melotte) and IC~1613, given that the Magellanic Clouds are too close to be effectively and efficiently used for this purpose. 
Other galaxies (NGC 2403, NGC 253, M81 and NGC 300), two to three times further away, are part of a calibration program being undertaken by McQuinn and her collaborators using both HST (Prop 15917) and JWST (Prop 1638)\footnote{ In that combined program McQuinn has obtained optical imaging data in F606W \& F814W filters using ACS, NIR imaging data in F110W \& F160W filters using WFC3-IR on HST. And, NIR imaging data in F090W, F150W \& F115W filters on both NIRCam and NIRISS, and MIR imaging data in F356W, F277W \& F444W filters on NIRCam, all using JWST.}.
\medskip\medskip\medskip\clearpage
\section{The Ground-based/Space-based Calibration}

Our fiducial bandpass/wavelength for undertaking this self-consistent calibration, is the ground-based I-band filter (centered at 8000\AA) and its space-based equivalent filters, F814W on ACS and F814W on WFC3-UVIS, both on HST (and conceivably the F090W filter on JWST/NIRCAM). The HST filters are very similar, with effective wavelengths of about 7970\AA ~(see Deustua \& Mack 2018) but, for the purposes of this paper, we will concentrate on examining the ground-based I-band data which has already been cross-matched to the rich sets of multi-wavelength observations reaching into the MIR, as provided by \textit{Spitzer} and the SAGE Survey (Meixner et al. 2006). The VIJHK data come from the Large Magellanic Cloud (LMC) stellar catalog published by Zaritsky et al. (2004) and cross-matched with 2MASS.  Zaritski et al. (2002) have compared their BVI photometry with the OGLE calibration (both of which are strictly on the Johnson-Cousins system) and conclude that their ``{\it average, global, photometric zero points are good to better than 0.03 mag}” and, more specifically, that  the individual offsets are 0.011, 0.038 and 0.002 mag for B,V and I, respectively\footnote{One caveat, noted by the referee, concerns the later revisions of the OGLE photometry, resulting from their introduction of ultra-precise interference filters for OGLE-IV, which subsequently fixed their OGLE-III photometry. OGLE-I and -II photometry may still be affected by long red tails in the original filters. However, the impact of these changes on this our results are far beyond the scope of this paper.}. We have adopted the I-band extinction of $A_I = $ 0.160~mag for the LMC TRGB stars, as discussed in detail in Freedman et al. (2019), which is consistent with the independent detailed study of Hoyt (2023)\footnote{
Lower values of the total line-of-sight reddening to the LMC have been reported by Skowron et al. (2021) and Nataf et al. (2021). The interested reader is referred to the latest discussion  of Red Clump extinction mapping found in Section 2 of  Hoyt (2021).}   Longer wavelength values were scaled using the Cardelli et al. (1989) extinction curve. However, we note that there are a number of alternative extrapolation available in the literature including Fitzpatrick \& Massa (2009), Fitzpatrick et al. (2019), Indebetouw et al. (2005), Schlafly \& Finkbeiner  (2011) (as given in NED galactic foreground extinction calculator) and most recently Newman et al. (2023). We use the differences in these various extrapolations to assign an uncertainty on our adopted values. For $A_J, A_H$ and $A_K$ we find externally calculated uncertainties of $\pm$0.015, 0.011 and 0.006~mag, respectively. In extrapolating our adopted total line-of-sight I-band extinction of 0.160~mag into the near-infrared those extinctions each carry a 15\% uncertainty.

\subsection{The Basic Equations}

We start by making explicit a simple but powerful point, which, in retrospect is obvious, but still needs to be stated.
The colors of TRGB stars, in whatever wavelength combinations that one cares to consider, can be expected to be linearly related, from band to band, over the small range of color defining the tip of the RGB population.

We adopt the $(J-K)_o$ color as fiducial for the most of this paper, while noting that any of other colors can be easily substituted once the distance-independent color-color transformations are defined.  

Figure 1 (left panel) shows the color-color transformation between $(V-I)_o$ and $(J-K)_o$ using LMC red giant branch (RGB) stars selected to fall within a $\pm$0.1 magnitude slice centered +0.3 mag below the flat trace of the I-band TRGB magnitude, as a function of color (see Figure 6). The right panel shows the same correlation run for stars centered on the +0.4 mag slice, +0.1 mag further, below the tip. No significant differences were found between the two solutions. We proceed with the sample defined by the brighter slice.
These stars significantly outnumber the number of stars defining the tip, but they all have colors that are sufficiently the same as the TRGB stars themselves and so they provide a much more (statistically) secure means of making the transformation of tip magnitude calibrations from one color to the next without
explicitly attempting to measure the individually more ill-defined slopes and zero points of new relations in any new color/magnitude combinations. 1550 RGB stars contributed to the solutions given here. They were constrained to fall within the color range $0.9 < (J-K) < 1.2 $ and $ 1.4 < (V-I) < 2.0$ mag.
The dispersion around the fit is $\pm$0.158 mag. Based on 1,550 stars, this gives an error on the mean of $\pm$0.004 mag. The fit in Figure 1 is 
$$(V-I)_o = 2.41~[(J-K)_o - 1.00] +1.61 ~(\pm 0.004)~~[1]$$

The three sub-panels in Figure 2 are color-color plots of the combinations of most immediate interest here: (J-K) vs (I-J), (J-K) vs (I-H) and (J-K) vs (I-K),top to bottom. 

$$(I-J)_o = 0.86~[(J-K)_o - 1.00] + 1.10 ~(\pm 0.02)~~[2]$$

$$(I-H)_o = 1.67~[(J-K)_o - 1.00] + 1.93 ~(\pm 0.04)~~[3]$$

$$(I-K)_o = 1.86~[(J-K)_o -1.00] +2.10 ~(\pm 0.02)~~[4]\footnote{Errors on the slopes are all at the $\pm$ 0.04 level}$$

\par\noindent
As given in Freedman (2021) and in the majority of other recent calibrations \footnote{Updating Table 1 in Freedman (2021) and centering all values at fiducial color of (V-I) = 1.5 mag, we have $M_I =$ -4.05~mag, Rizzi et al. (2007); -4.05~mag, Bellazzini et al. 2008; -4.05~mag, Tammann et al. 2008; -4.05~mag, Madore et al. 2009; -4.01~mag, Jang \& Lee 2017; Reid et al. 2019; -3.97~mag, Yuan et al, 2019, reddened; -4.05~mag, Freedman et al 2020; -4.05~mag, Hoyt et al. 2021, LMC; -4.05~mag, Hoyt et al. 2021, SMC. {\bf To be complete, it can be mentioned that Soltis et al. 2021 obtained a value of $M_I = $ -3.97$\pm$0.06~mag, based on a Gaia parallax to $\omega$ Cen, the uncertainty of which has been challenged by Vasiliev \& Baumgardt 2021.}} $$M(I)_o = -4.05$$~mag  it then  
immediately follows that 
$$M(V)_o = -4.05 + 1.00~(V-I)_o~~[5]$$
and then using Equation 1
$$M(V)_o = -2.44  + 2.41~[(J-K)_o -1.00]~~[6]$$ 
after transforming to $(J-K)_o$ and re-centering the color.
It then also follows that

$$[M(I)_o - M(J)_o] = 0.86~[(J-K)_o - 1.00] + 1.10~~[7]$$

$$[M(I)_o - M(H)_o] = 1.67~[(J-K)_o -1.00] + 1.93~~[8] $$

$$[M(I)_o - M(K)_o] = 1.86~[(J-K)_o - 1.00] + 2.10~~[9] $$
\medskip\par
Again, by substituting $M(I)_o = -4.05$ into the above three equations, collecting the respective zero points and re-centering the equations at $(J-K)_o = 1.00$ gives 

$$M(J)_o = -5.15 - 0.86~[(J-K)_o - 1.00]~~[10]$$

$$M(H)_o = -5.98 -1.67~[(J-K)_o - 1.00]~~[11]$$

$$M(K)_o = -6.15 -1.86~[(J-K)_o - 1.00]~~[12]$$

\par\noindent
Similarly, for the mid-infrared absolute magnitudes we derive the following relations,

$$M(3.6)_o = -6.29 -2.30~[(J-K)_o - 1.00]~~[13]$$
 
$$M(4.5)_o = -6.18 -2.30~[(J-K)_o - 1.00]~~[14]$$

\noindent
and

$$M(J)_o = -5.15 -0.72~[(J-3.6)_o - 1.20]~~[15]$$


\noindent
After adjusting for the slightly (5\%) shorter central wavelength of the JWST F115W filter, and the longer color baseline of [(F115W -3.6] compared to (J-K), we get a first estimate of 

$$M(F115W)_o = -5.01 -0.72~[(F115W-3.6)_o - 1.34]~~[16]$$

\noindent
with the formal error on the color term being $\pm$0.06. This slope of -0.72 is well within the range of values (weighted least squares: -0.74 $\pm$ 0.17; unweighted least squares: -0.70 $\pm$ 0.16; theory for a 10 Gyr population: -0.74 $\pm$ 0.03) quoted by Hoyt et al. (2023) for the JWST observations of the TRGB found in the galaxy NGC~4536, where the color baseline used F444W which is largely degenerate with 3.6 microns for these stars.

\begin{figure}
\centering
\includegraphics[width=16.0cm, angle=-00]{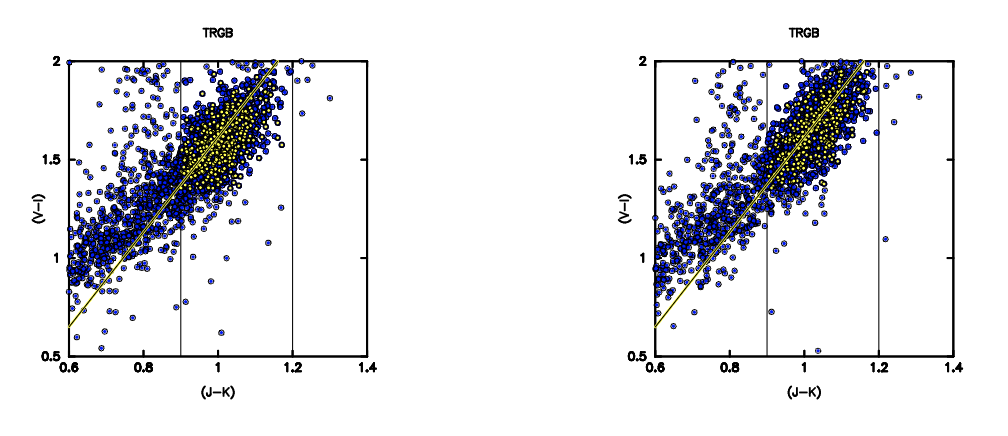}
\caption{(V-I) vs (J-K) Color-Color Plots. Left Panel: Stars are selected to be 
+0.3 ($\pm$ 0.1) mag below the I-band TRGB. Right Panel: Stars are chosen to be +0.4 ($\pm$ 0.1) mag below the TRGB. No significant difference is found between these two solutions for color-color correlation used in Equation [1]. } 
\end{figure}

\begin{figure}
\centering
\includegraphics[width=8.0cm, angle=-00]{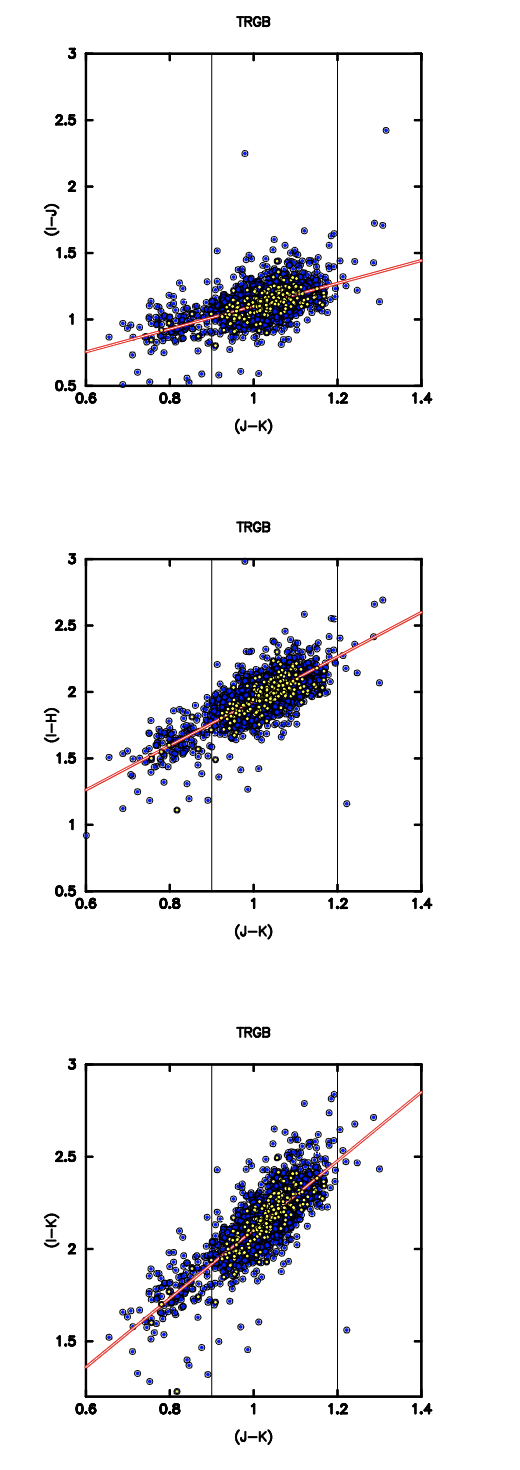}
\caption{Color-Color Transformation Plots: Upper plot: (I-J) to (J-K), middle plot: (I-H) to (J-K), and lower plot: (I-K) to (J-K). Solid red lines are the adopted linear fits to the data.  Yellow points represent one in five of the blue data points and are over-plotted to give a visual sense of the surface density of points in high-overlap regions.}
\end{figure}

\medskip\medskip
\begin{figure}
\centering
\includegraphics[width=14.5cm, angle=-00]{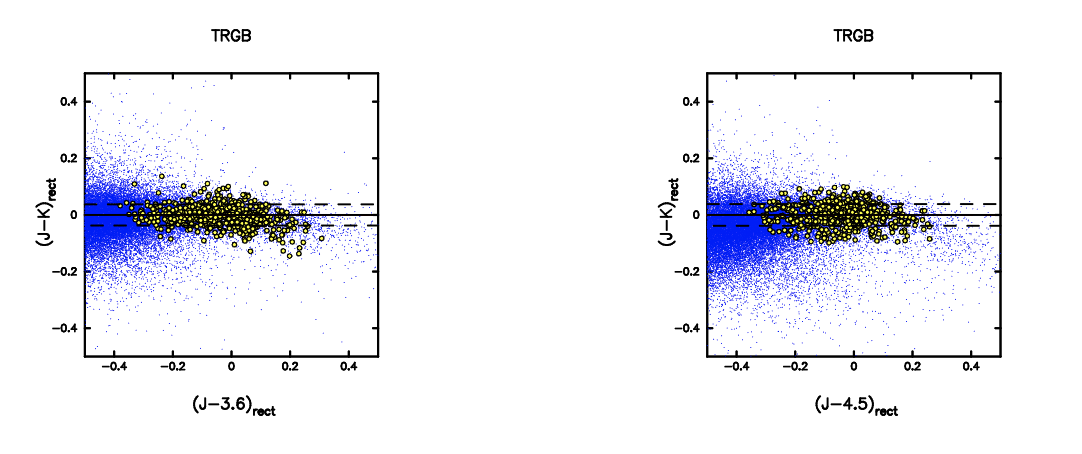}
\caption{Rotated Color-Color Plots for (J-K) versus the Mid-Infrared blended colors: (J-3.6) and (J-4.5) in the left and right panels, respectively. See text for additional details.}
\end{figure}

\begin{figure}
\centering
\includegraphics[width=9.0cm, angle=-00]{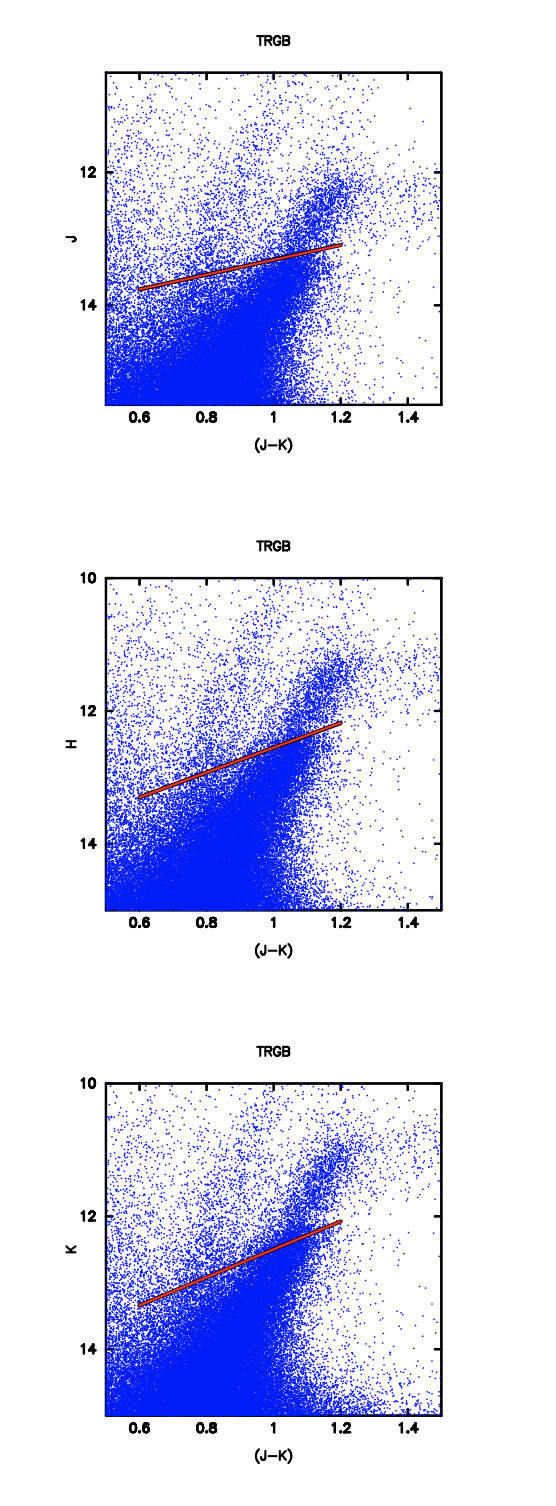}
\caption{NIR Color-Magnitude Diagrams centered on the TRGB in J, H and K versus (J-K), top to bottom.
The red slanting line is defined by the color-color transformations shown in three panels of Figure 2. }
\end{figure}

\clearpage
\begin{figure}
\centering
\includegraphics[width=1.2\textwidth]{"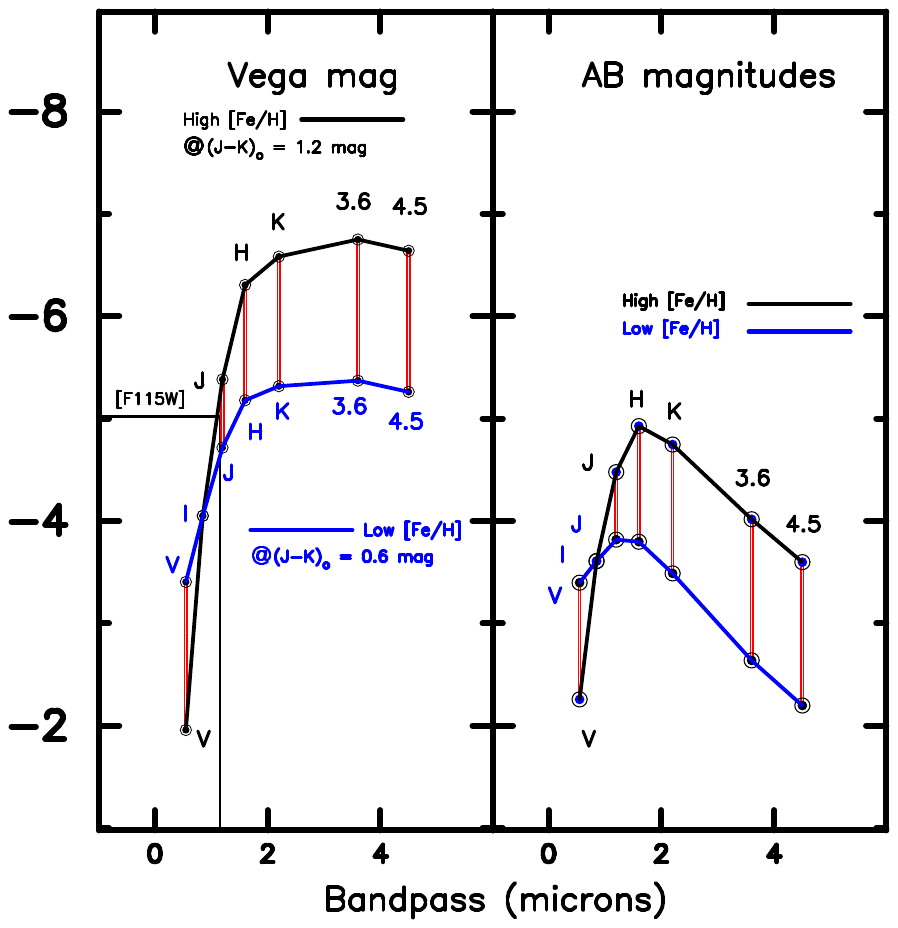"}
\caption{The trends, with increasing wavelength (bandpass) of the absolute magnitude of TRGB stars at the blue [(J-K) = 0.6 mag] metal-poor end and at the red [(J-K) = 1.2 mag] metal-rich end of the fits shown in Figure 6. The thin black lines indicate where the F115W filter on JWST/NIRCam is expected to show its slanting tip 
which because of that band's slightly blue effective wavelength than ground-based J, it will be slightly fainter over-all and slightly shallower in its slope than the J-band values shown here and elsewhere. The JWST absolute mean-magnitude zero point is set at the fiducial  color [F115W - F365W] = 1.34~mag, corresponding to (J-K) = 1.00~mag. }
\end{figure}

\section{ Conclusions}
Figure 6 graphically summarizes the results of this paper. The black circled points map out $I_o$ vs $(V-I)_o$ color-magnitude diagram for stars in the LMC. One star in 10 is colored yellow (and then one star in 5 is colored red) so as to give some sense of the density of points in heavily crowded regions of the diagram. 
Large yellow points mark the stars (found below the flat I-band TRGB) that are used in all of the subsequent color-color transformations. 
Their mean level is indicated by the horizontal red line. The fiducial I-band TRGB is shown by the horizontal blue line, parallel to and 0.3 mag above the red line. 
The positively-sloped lines above the I-band TRGB are the NIR and MIR tracers of the TRGB in the JHK NIR bands, followed by the (overlapping) 3.6 and 4.5$\mu$m MIR TRGB relations. 
These lines are at their magnitudes in their respective filters (still plotted against the common $(J-K)$ color. The downward sloping blue line, well below the I-band, is the $V_o$ vs $(J-K)_o$ TRGB relation; it has a slope of unity. 
The dramatic increase in luminosity of the TRGB stars seen in the longer-wavelength filters is impressive, but it needs to be pointed out that this advantage saturates in the MIR where the long-wavelength colors become zero on the Raleigh-Jeans portion of the spectral energy distribution.

\begin{figure*}
\centering
\label{fig:f5}
\includegraphics[width=0.9\textwidth]{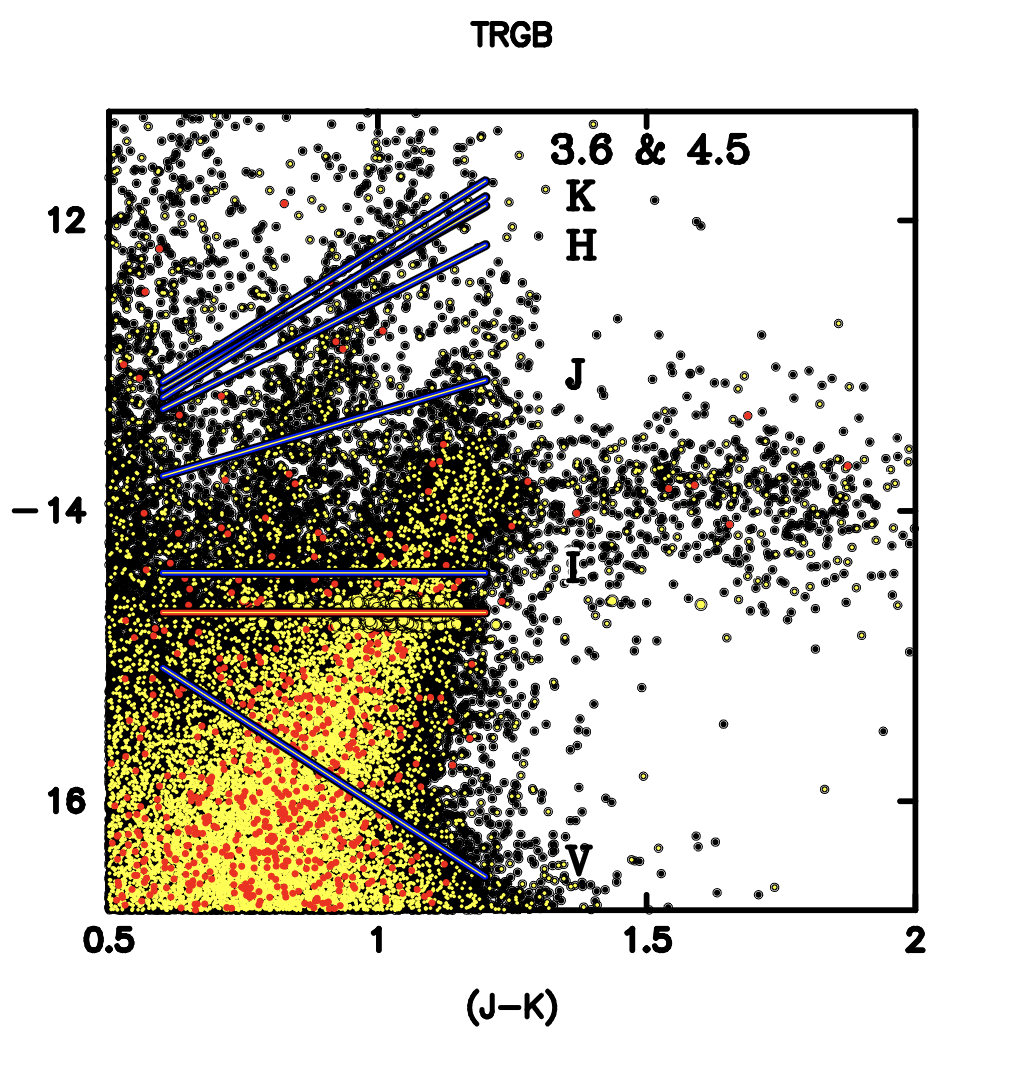}
\caption{The individual traces (in blue) of the tip of the red giant branch in VIJHK and in 3.6 \& 4.5 microns (the latter two having identical slopes). The TRGB I-band slope is flat blue line. 
Below it is a red line, +0.3 mag fainter, going through larger yellow dots that are the tracer stars used to define the color-color plots shown in this paper.}
\end{figure*}

\medskip\medskip

\begin{deluxetable}{l|cc|cc}
\tablecaption{LMC Vega magnitudes and LMC AB magnitudes of Multi-Wavelength Tip Calibrations. Blue ends of the tip in Columns 2 and 3; Red ends in Columns 4 and 5.  \label{ABmags}}
    
\tablehead{\colhead{Filter} & \colhead{Vega (B)} & \colhead{AB(B)}& \colhead{Vega(R)} & \colhead{AB(R)}}

\startdata
$V$ & $15.08$ & $15.08$ & $16.22$ & $16.22$  \\
$I$ & $14.43$ & $14.87$ & $14.43$ & $14.87$ \\
$J$ & $13.76$ & $14.66$ & $13.10$ & $14.00$ \\
$H$ & $13.30$ & $14.68$ & $12.17$ & $13.55$ \\
$K_{s}$ & $13.16$ & $14.99$ & $11.90$ & $13.73$ \\
$[3.6]$ & $13.11$ & $15.84$ & $11.73$ & $14.46$ \\
$[4.5]$ & $13.22$ & $16.28$ & $11.84$ & $14.88$ \\
\enddata
\end{deluxetable}

\vfill\eject
\section{Acknowledgements}  
We thank the {\it University of Chicago} 
and {\it Observatories of the Carnegie Institution for Science} for their support of our long-term research into the calibration and determination of the expansion rate of the Universe. 
Support for this work was also provided in part by NASA through grant number HST-GO-13691.003-A from the Space Telescope Science Institute, which is operated by AURA, Inc., under NASA contract NAS~5-26555. 

We thank both Margaret Meixner and Dennis Zaritsky for making their individual catalogs of resolved-star photometry of the LMC publicly available. We also thank the referee for suggesting that we investigate the effects of alpha-enhanced metallicities on the level of the TRGB and on the slopes in the near-infrared.
\clearpage
 APPENDIX A
\medskip\medskip
\par {BASTI Models}
\medskip\medskip
\par

 We have accessed the  stellar evolutionary models provide by Cassisi \& Salaris available at their website (http://albione.oa-teramo.inaf.it/) as described in Pietrinferni et al. (2004, 2006). Six isochrones for two sets of models were downloaded. One adopted a standard solar composition; the other was enhanced in alpha-elements. Each of the six models (shown in Figure 7) had the following metallicities: Z = 0.0001, 0.003, 0.006, 0.010, 0.020 and 0.040. As is well known the terminal points of the RGB stars (in color) are a monotonic, increasing function of increasing metallicity. This is demonstrated in both the alpha-enhanced model (right panel) and somewhat more dramatically in the solar mixture models (left panel). 

 The effect of alpha enhancement on the absolute magnitude of the TRGB is small, and is highlighted by the two horizontal lines duplicated in each of the two panels.
The dashed line represents 
the approximate level of the TRGB averaged over the three brightest tips corresponding to metallicities of Z = 0.003, 0.008 and 0.010.
That TRGB level is $M_I =$ -4.037~mag for the solar mixture model, and $M_I =$ -4.050~mag for the alpha-enhanced mode.\footnote{Note that we have offset the two BASTI magnitude scales by +0.186~mag so as to have the alpha-enhanced TRGB shown in Figure 8 to have an absolute magnitude of $M_I = $ -4.05~mag, simply to make the CMDs correspond more closely to the generally adopted empirical zero point.}

The downward sloping arrow attached to the TRGB of the lowest metallicity isochrone in each of the plots is the full range of the color and luminosity effect due to age shown in
the accompanying Figure 8 (below). If, in a given galaxy halo, there are a range of ages the net result would be to have the oldest population always defining the observed TRGB with more recent populations simply adding counts to the fainter extension of the summed RGB luminosity function. Assuming that every halo has first-generation, low-mass, low-metallicity stars that enrich the gas with metals
for subsequent generations of halo stars, then they will be the same stars in all galaxies determining their TRGB distances. Later (merger) events would only contribute fainter RGB stars and not impact that fact. 

If there were a compelling reason to suspect that galaxies had varying amounts of populations of RGB stars differing randomly in their solar versus alpha-enhanced elemental mixtures, then the range in the magnitude of the TRGB seen here ($\Delta M_I $ = 0.013~mag), expressed as a dispersion, would effectively blur the tip by only $\pm$0.004~mag. 

 For readers who are not familiar with the effects of age on the brightness and color of the TRGB, we have included Figure 8, which we note has a highly magnified color range emphasizing the small (but still systematic) effect of age.  To emphasize this point, the full effect of age as given in Figure 8 is also shown in Figure 7 as a downward-pointing arrow attached to the terminal point of the lowest-metallicity isochrone in the two panels, with younger populations falling systematically below the oldest stars of all metallicities. In other words, if there are predominantly old halos with a contribution of younger stars (produced by whatever mechanism) their presence will not affect the detected magnitude of the TRGB.
\begin{figure}
\centering
\includegraphics[width=0.4\textwidth]{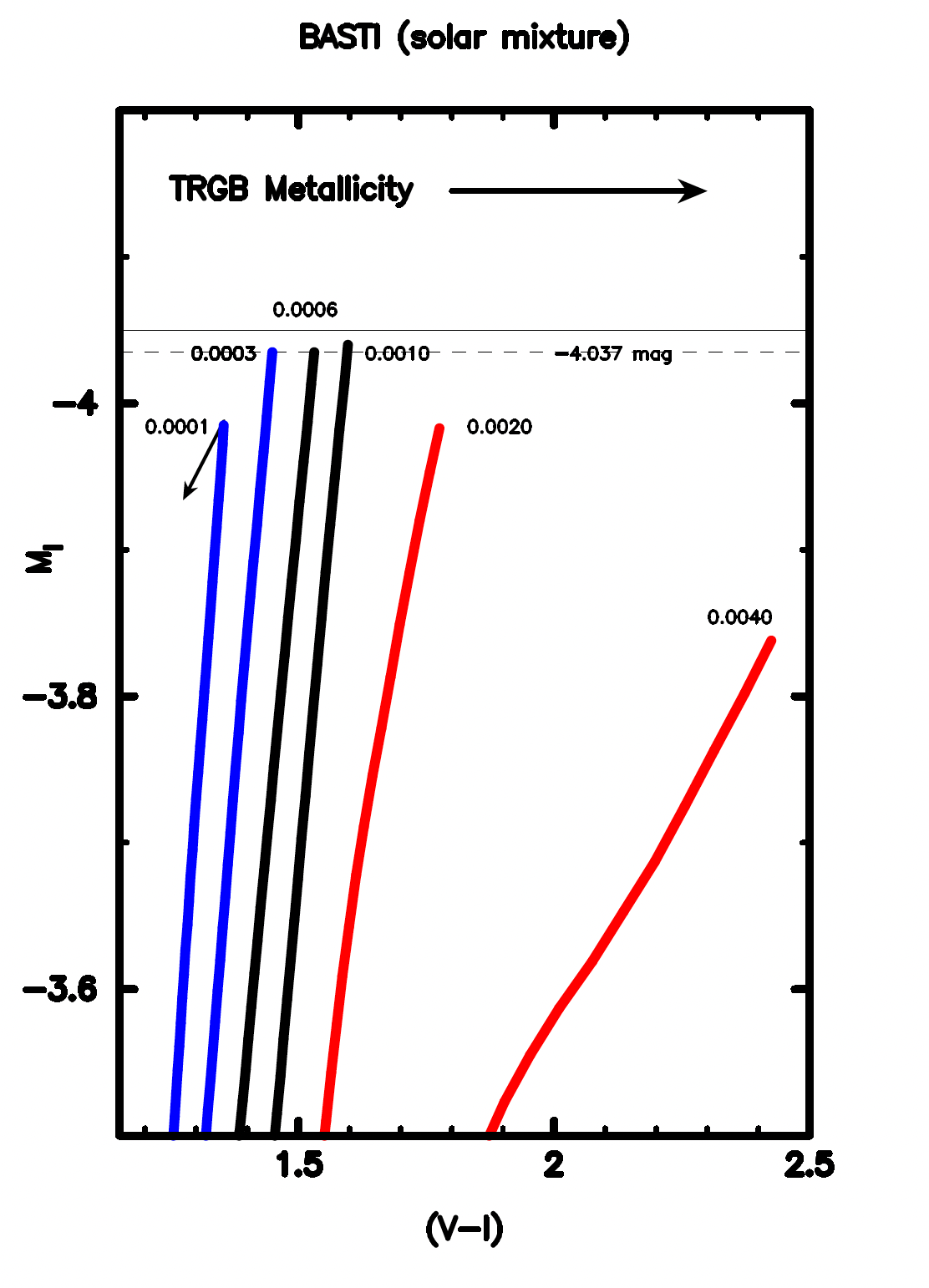}
\includegraphics[width=0.4\textwidth]{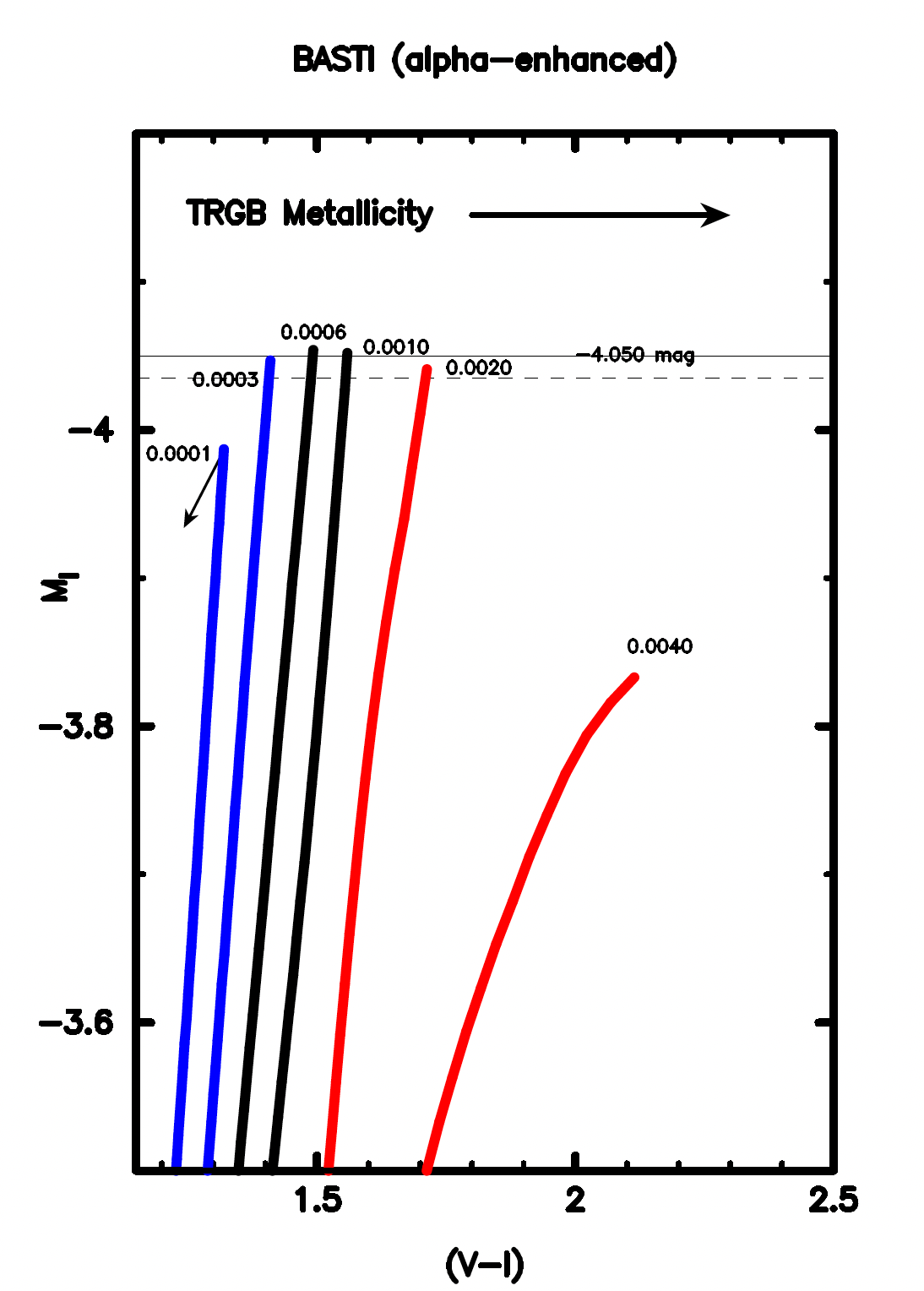}
\caption{The trends of RGB isochrones with metallicity for alpha-enhanced models (in the right panel) and solar models (in the left panel). The horizontal dashed lines are at the approximate level of the brightest TRGB stars (at -4.037~mag) having a solar mixture; the solid horizontal line marks the level of the TRGB (at -4.050~mag) for alpha-enhanced models. See text for additional details.}
\end{figure}

\begin{figure}
\centering
\includegraphics[width=0.7\textwidth]{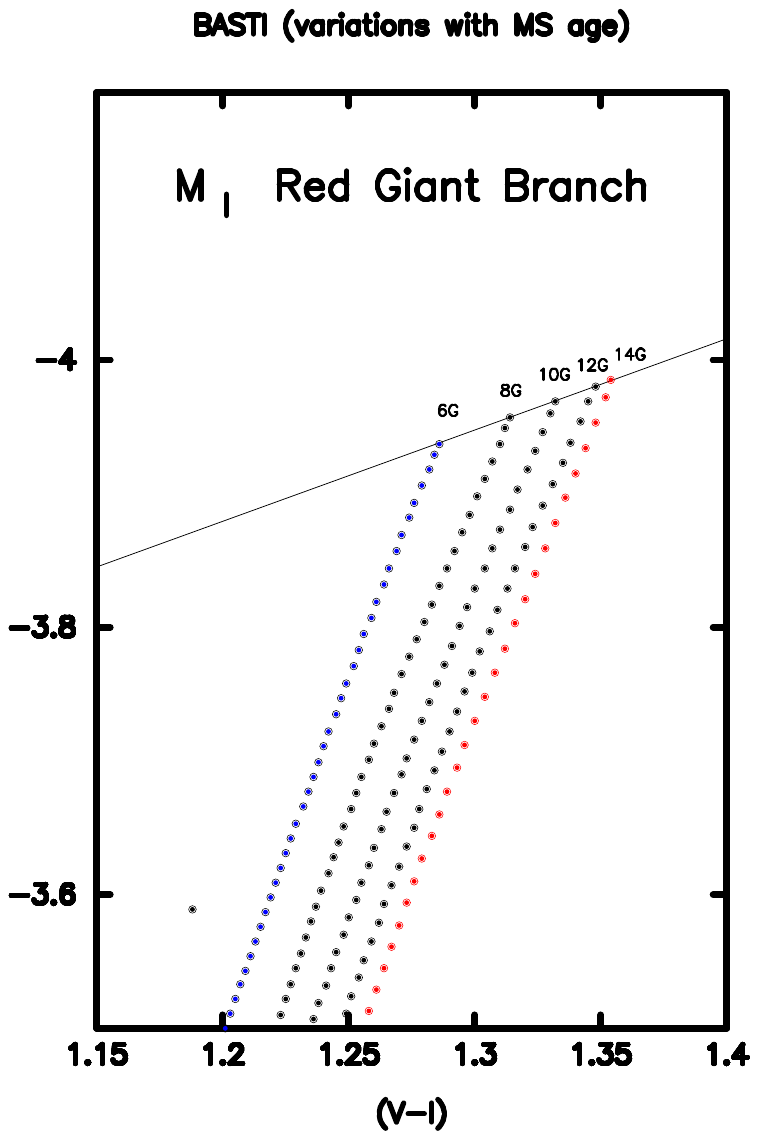}
\caption{BASTI Models of the the Red Giant Branches of stellar populations having the same solar abundances  but differing in ages, ranging from 6~Gyr (blue points, to the left) up to 14 Gyr (red points, to the right). Note that the color range spanned by these models at the TRGB is very small (0.08 mag) in (V-I) with a trend toward bluer colors and fainter TRGB magnitudes with decreasing age. The full run of color and magnitude with age is shown as a downward slanting arrow in Figure 7 (left and right panels). }
\end{figure}

\begin{figure}
\centering
\includegraphics[width=0.4\textwidth]{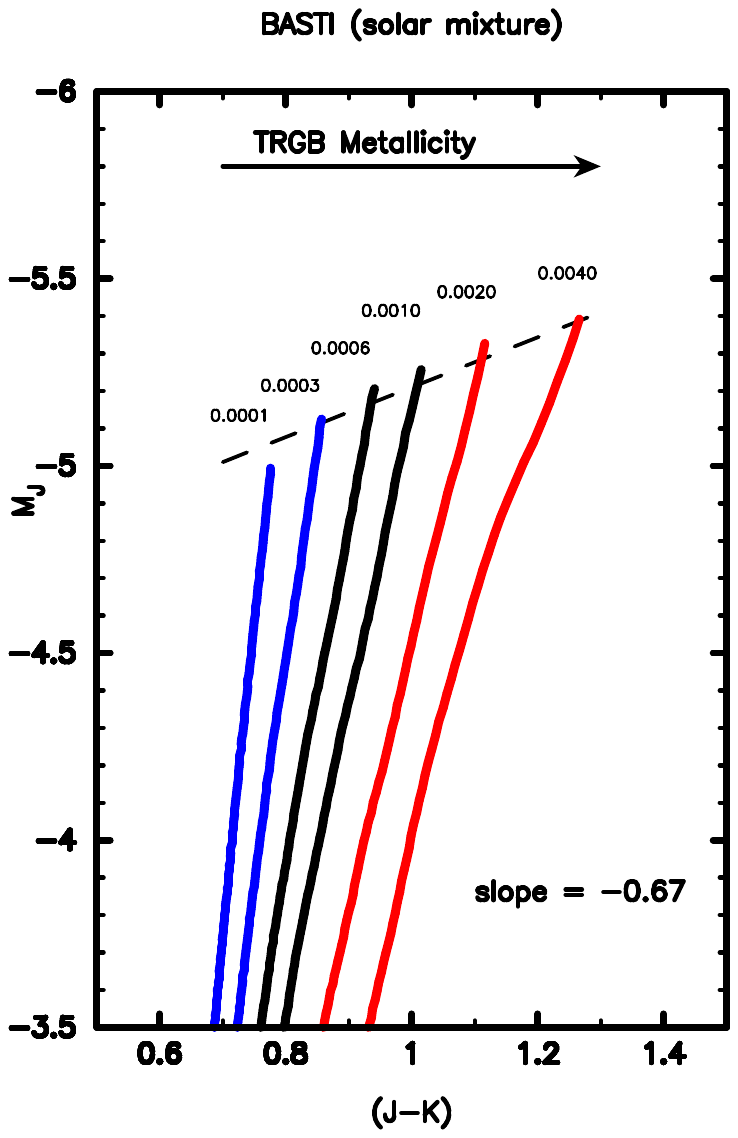}
\includegraphics[width=0.4\textwidth]{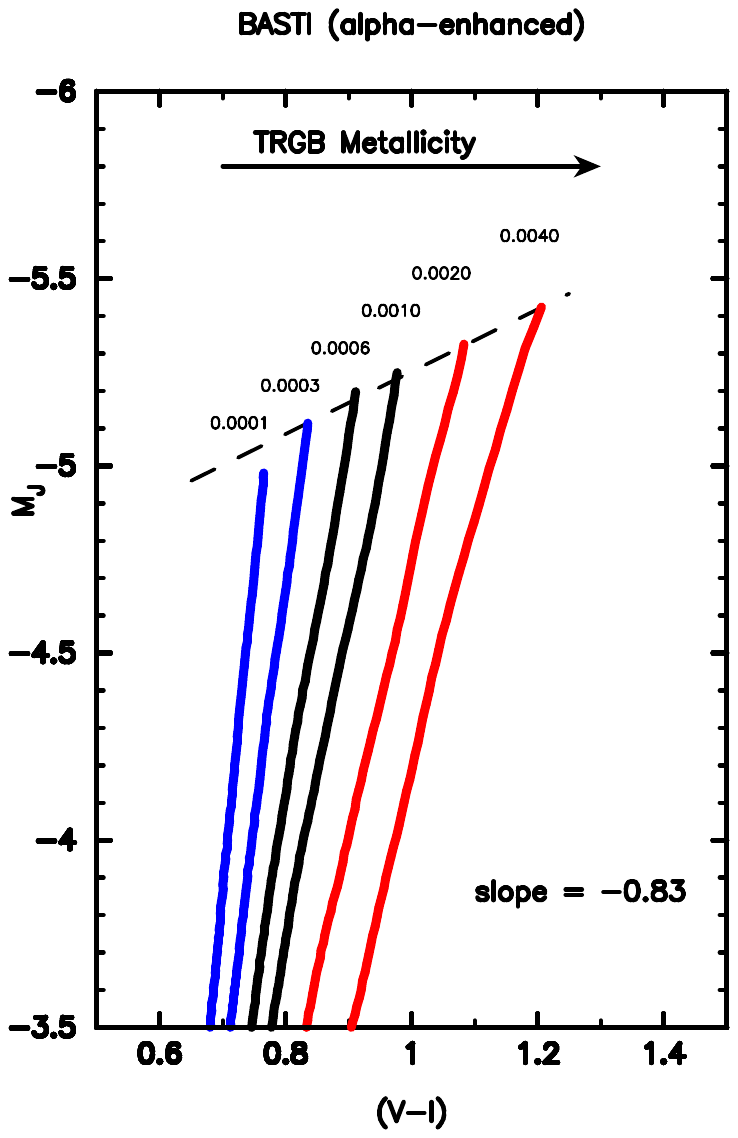}
\includegraphics[width=0.4\textwidth]{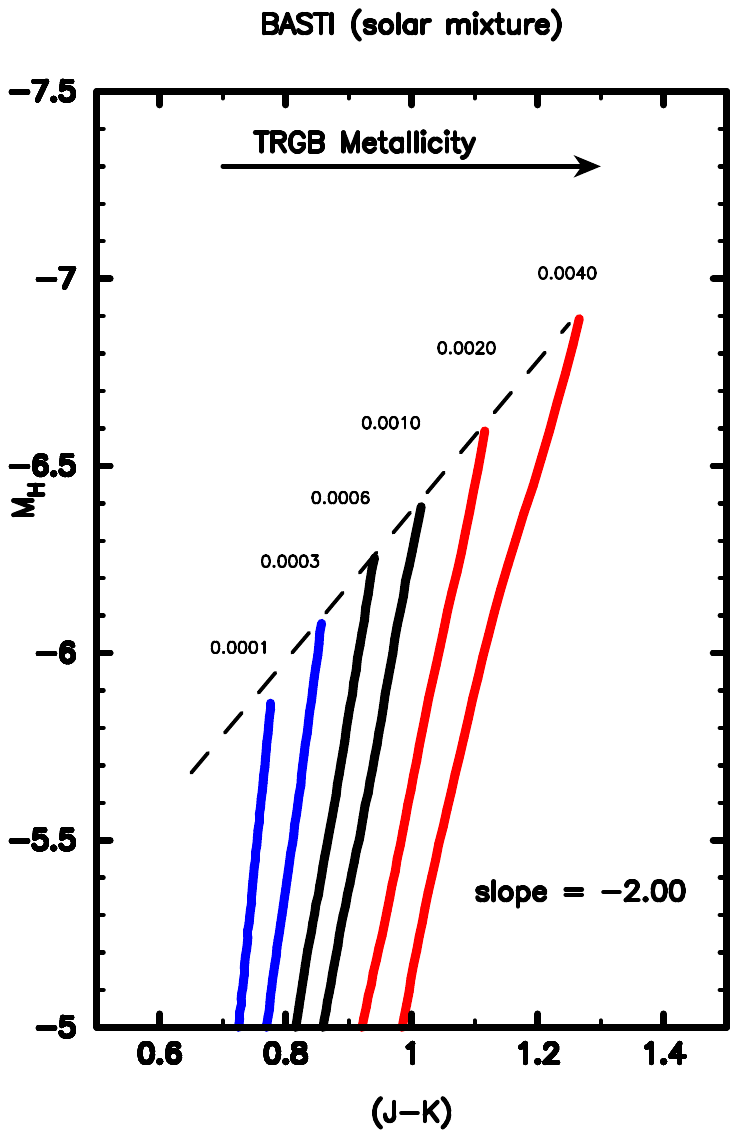}
\includegraphics[width=0.4\textwidth]{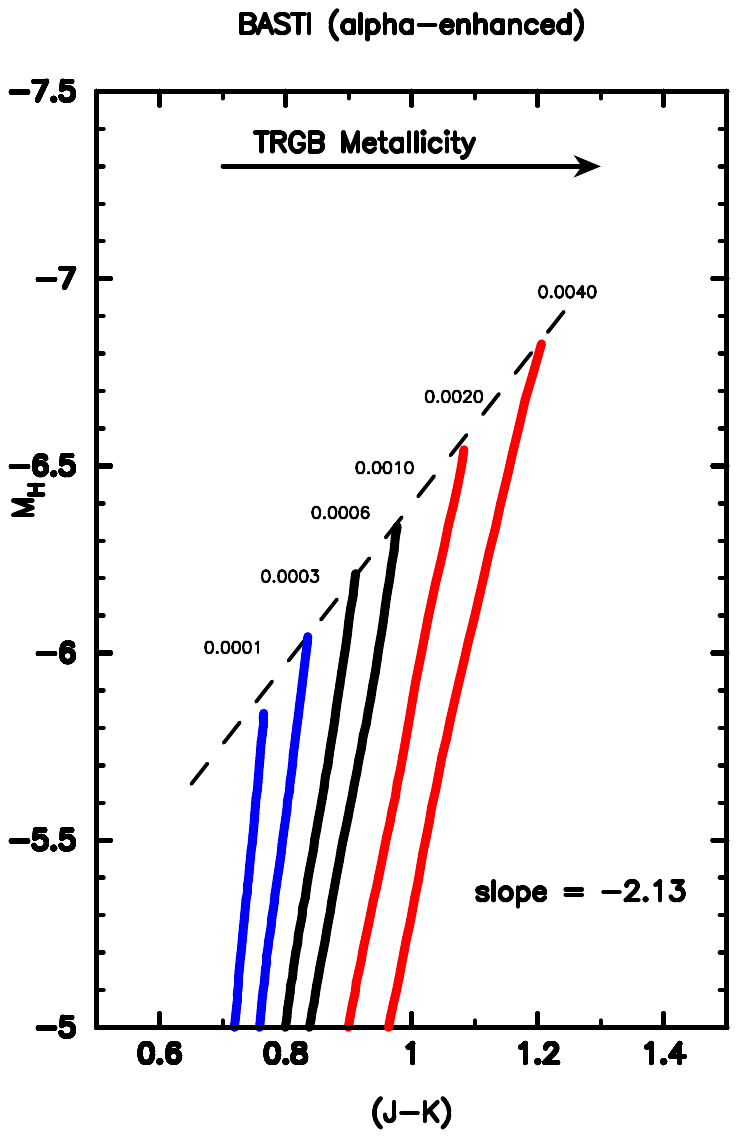}
\includegraphics[width=0.4\textwidth]{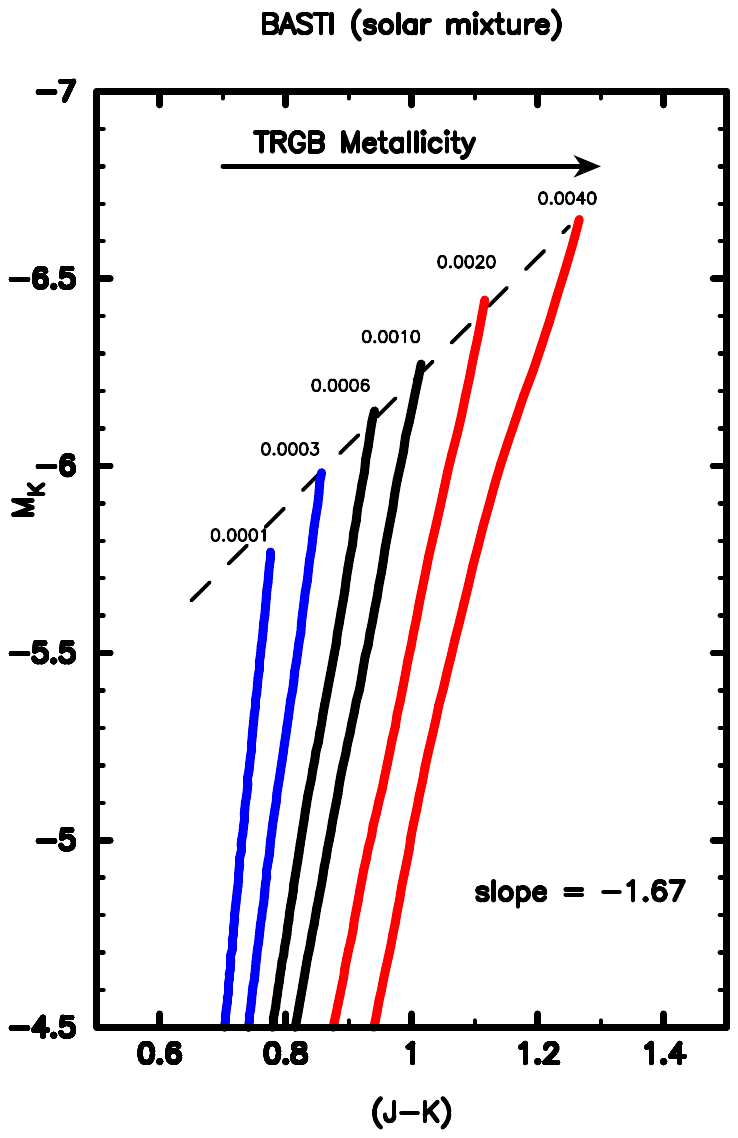}
\includegraphics[width=0.4\textwidth]{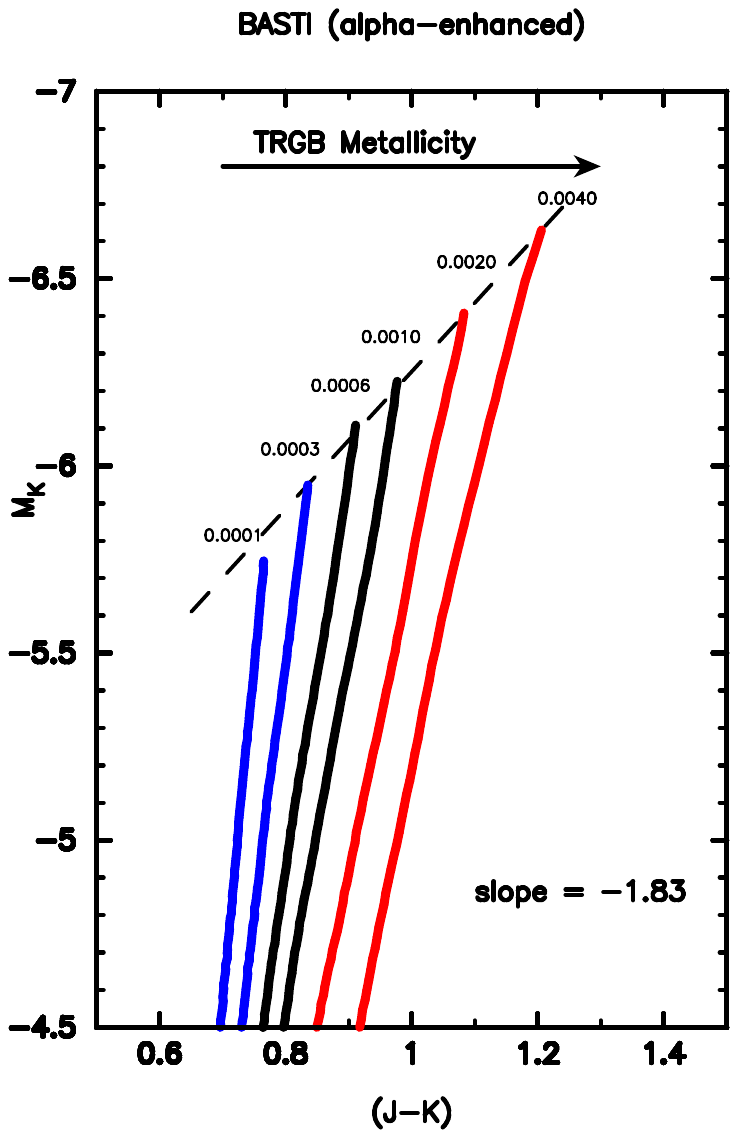}
\caption{Near-Infrared (JHK) Basti models of red giant branch isochrones for a variety of metallicities (in each panel) and for a solar mixture in the three (JHK, top to bottom) panels in left column, and an alpha-enhanced mixture in the three (JHK) panels in the right column. The dashed black line is a fit to the TRGB as defined by the five highest metallicity isochrones. Those slopes are individually given in the lower right portion of each of the color-magnitude diagrams.}
\end{figure}
Finally, we use the same set of BASTI models to examine the effects of alpha-element enhancements on the slopes of the tips of the red giant branches in the three near-infrared JH \& K bandpasses. A mosaic of those color-magnitude diagrams is found in Figure 9. The left column of plots are for a solar metallicity, running from J vs (J-K) at the top, followed by H vs (J-K) and then K vs (J-K) at the bottom.  The right column follows the same sequence, but it is for alpha-enhanced models. Slopes (fit by dashed black lines to the five highest metallicity isochrones) are given to the lower right of each of the panels.

The main conclusions to be gleaned from these plots are as follows: (a) the solar models are more spread out to the red than the alpha-enhanced models, but (b) they have slightly larger slopes in J (-0.67 vs -0.83) and K (-1.67 vs -1.83)\footnote{Note however that the J and K slope solutions are not independent, but it is easily shown that Slope(K) = Slope(J) -1.0 is true by definition.}, and a smaller slope in H (-2.13 vs -2.00) for the alpha-enhance models. These slopes can be compared with the LMC slopes (-0.86, -1.67  \& -1.86) given above (Section 2.1, Equations 10, 11, and 12) and the other set of empirical slopes given by Madore et al. (2018) for the NIR TRGB data of stars in IC~1613 and the LMC (Hoyt et al. 2018).  Those slopes are -0.85, -1.62 and -1.85 in JK \& K, respectively. They, in turn, are closely matched by the self-consistent J and K slopes (-0.811 and -1.811, respectively) theoretically modelled by Serenelli et al. (2017) for the color range 0.76 $<$ 1.50~mag.  These values may be suggesting that the alpha-enhanced models are more representative of the halo stars in the LMC and IC~1613. How this geneneralizes out to other halos will have to await high-precision slopes being independently measured in more galaxy halos. 
\clearpage
\section{References}
\medskip
\par\noindent
Bellazzini, M. 2008, MmSAI, 79, 440
\par\noindent
Cardelli, J.A., Clayton, G.C., Mathis, J.S. 1989, ApJ, 345, 245
\par\noindent
Deustua, S.E. \& Mack, J. 2018, {\it Comparing the ACS/WFC and WFC3/UVIS Calibration and Photometry}, 

HST Instrument Science Reports: ACS-2017-02 and WFC3-2018-02

\par\noindent
Fitzpatrick, E.L., \& Massa, D., Gordon, K.D. 2009, ApJ, 886, 108

\par\noindent
Fitzpatrick, E.L., \& Massa,  2019, ApJ, 699, 1209

\par\noindent
Freedman, W.L., Madore, B.F., Hatt, D. 2019, ApJ, 882, 24 arXiv: 1907.05922

\par\noindent
Freedman, W.L., Madore, B.F., Hoyt, T., et al. 2020, ApJ, 891, 57

\par\noindent
Freedman, W.L. 2021, ApJ, 919, 16 arXiv: 2106.15656

\par\noindent
Hoyt,T.J., Beaton, R.L., Freedman, W.L., et al. 2021, ApJ, 915, 34

\par\noindent
Hoyt, T.J.
2023, NatAs, 7, 590

\par\noindent
Hoyt, T.J., Freedman, W.L., Madore, B.F., et al. 2018, ApJ, 858, 12

\par\noindent
Hoyt, T.J., Jang, I.-S., Freedman, W.L. et al. 2023, ApJ, (submitted).

\par\noindent
Indebetouw, R., Mathis, J.S., Babler, B.L. et al. 2005, ApJ, 619, 931 

\par\noindent
Jang, I.S., \& Lee, M.G. 2017, ApJ, 835, 28

\par\noindent
Lee, A. J., Freedman, W. L., Madore, B. F., et al., 2021, ApJ, 907, 112.

\par\noindent
Madore, B.F., Mager,V., \& Freedman, W.L. 2009, ApJ, 690, 389

\par\noindent
Madore, B.F., Freedman, W.L, Hatt, D., et al. 2018, ApJ, 858, 11

\par \noindent
Madore, B.F., \& Freedman, W.L. 2020, AJ, 160, 170 

\par\noindent
Newman, M.J.B., McQuinn, K.B.W., Skillman, E.D. 2023, ApJ, submitted

\par \noindent
Pietrzynski, G., Graczyk, D., Gallenne, A., et al. 2019, Nature, 567, 200
\par\noindent
Pietrinferni, A., Cassisi, S.,  Salaris, M. \&  Castelli, F. 2004, ApJ, 612, 168
\par\noindent
Pietrinferni, A., Cassisi, S.,  Salaris, M. \&  Castelli, F. 2006, ApJ, 642, 797
\par\noindent
Planck Collaboration, Aghanim, N., Akrami, Y., et al. 2020, \aap, 641, A6. 
\par\noindent
Reid, M.J., Pesce, D.W., \& Riess, A.G. 2019, ApJL, 886, L27
\par\noindent
Rizzi, L., Tully, R.B., Makarov, D., et al. 2007, ApJ, 661, 815
\par\noindent
Schlafly, E.F., \& Finkbeiner, D.P. 2011, ApJ, 737, 103 
\par\noindent
Serenelli, A., Weiss, A., Cassisi, S.  2017, A\&A, 606, A33 

\par\noindent
Tammann, G.A., Sandage, A., \& Reindl, B. 2008, ApJ, 679, 52

\par\noindent
Vasiliev, E., \& Baumgardt, H. 2021, MNRAS, 505, 5978

\par\noindent
Yuan,W., Riess, A.G., Macri, L.M., et al. 2019, ApJ, 886, 61

\par\noindent
Zaritsky, D., Harris, J., Thompson, I.B., et al. 2004, AJ,  128, 1606
\vfill\eject

\end{document}